%% file: ms.tex
\documentclass[aps,prd,twocolumn,groupedaddress,superscriptaddress]{revtex4-2}

\usepackage{lipsum}
\usepackage{graphicx}  
\usepackage{dcolumn}   
\usepackage{bm}        
\usepackage{amssymb}   
\usepackage{amsmath}
\usepackage{xcolor}
\usepackage{hyperref}

\hypersetup{colorlinks=true, linkcolor=blue, citecolor=blue, urlcolor=blue}

\begin{document}


\title{\bf Search for invisible axion dark matter of mass m$_a=43~\mu$eV with the QUAX--$a\gamma$ experiment}

\input{authors.tex}

\date{\today}

\begin{abstract}
A haloscope of the QUAX--$a\gamma$ experiment composed of an oxygen-free high thermal conductivity-Cu cavity inside an 8.1~T magnet and cooled to $\sim200$~mK is put in operation for the search of galactic axion with mass $m_a\simeq43~\mu\text{eV}$. The power emitted by the resonant cavity is amplified with a Josephson parametric amplifier whose noise fluctuations are at the standard quantum limit. With the data collected in about 1~h at the cavity frequency $\nu_c=10.40176$~GHz, the experiment reaches the sensitivity necessary for the detection of galactic QCD-axion, setting the $90\%$ confidence level limit to the axion-photon coupling $g_{a\gamma\gamma}<0.766\times10^{-13}$ GeV$^{-1}$.
\end{abstract}


\maketitle

\section{\label{sec:intro}Introduction}

The axion is a hypothetical particle that was first introduced by Weinberg~\cite{Weinberg} and Wilczek~\cite{Wilczek} as a consequence of the Peccei-Quinn mechanism to solve the strong $CP$ problem of QCD~\cite{PecceiQuinn,*PecceiQuinn2}. The axion is a pseudo-Goldstone boson associated with an additional symmetry of the Standard Model Lagrangian, which is spontaneously broken at an extremely high energy scale $f_a$. Axions, with mass $m_a \propto 1/f_a$, may constitute the dark matter (DM) content in our Galaxy~\cite{preskill1983,*abbott1983,*dine1983}. Astrophysical observations and cosmological considerations suggest a favored mass range of $1~\mu\text{eV} < m_a < 10~\text{meV}$~\cite{IrastorzaRedondo,*PDG2018}.
Several operating and proposed experiments rely on the haloscope concept proposed by  Sikivie~\cite{Sikivie,*Sikivie2} to probe the axion existence; among them are ADMX~\cite{admxPRL120,admxPRL124}, HAYSTAC~\cite{HAYSTAC3}, ORGAN~\cite{ORGAN}, CAPP-8T~\cite{CAPP8T}, CAPP-9T~\cite{CAPP9T}, RADES~\cite{RADES,*RADES2},  QUAX~\cite{QUAX,QUAX2,QUAX4,QUAX3}, and KLASH~\cite{KLASH,*KLASH2,*KLASH3}. Dielectric haloscopes have also been proposed, like MADMAX~\cite{MADMAX} and BRASS~\cite{BRASS}.

The classical haloscope detection scheme consists of a resonant cavity immersed in a static magnetic field to stimulate the axion conversion into photons through the Primakoff effect. When the cavity resonant frequency $\nu_\text{c}$ is tuned to the axion mass $m_ac^2/h$, the expected power deposited by DM axions is given by~\cite{HAYSTAC,*HAYSTAC2}
\begin{equation}
	\label{eq:power}
	P_{a}=\left( \frac{g_{a\gamma\gamma}^2}{m_a^2}\, \hbar^3 c^3\rho_a \right) \times
	\left( \frac{\beta}{1+\beta} \omega_c \frac{1}{\mu_0} B_0^2 V C_{010} Q_L \right),
	\end{equation}
where $\rho_a \sim 0.4 - 0.45$\,GeV/cm$^3$~\cite{pdg-axions} is the local DM density and $g_{a\gamma\gamma}$ is the coupling constant appearing in the Lagrangian describing the axion-photon interaction.
The second set of parentheses contains the vacuum permeability $\mu_0$, the magnetic field strength $B_0$, the cavity volume $V$, its angular frequency $\omega_c=2\pi\nu_c$, the coupling between the cavity and receiver $\beta$, and the loaded quality factor $Q_L=Q_0/(1+\beta)$, with $Q_0$ the unloaded quality factor. Here, $C_{010}$ is a geometrical factor equal to about 0.69 for the $TM$010 mode of a cylindrical cavity. When the cavity is not exactly tuned to the axion mass, the Lorentzian behavior needs to be taken into account, so Eq.~(\ref{eq:power}) is multiplied by
\begin{equation}
	\label{eq:lorentz}
	\frac{1}{1+\left(2Q_L \, \delta \omega / \omega_c \right)^2},
	\end{equation}
where $\delta \omega$ is the detuning from resonance.

Presently, different solutions are being devised to improve the signal-to-noise ratio. The resonant cavity design is moving towards the multiple-cell concept~\cite{CAPP9T} and the employment of different materials, like superconductors (Refs.~\cite{QUAX3,IEEEquax} and Ref.~\cite{ybcoCav}) or dielectrics~\cite{PhotCav,PhotCavNIM}. On the amplification side, state-of-the-art experiments operate at the SQL $-$ with SQUIDs~\cite{admxPRL120,SQUID-axions} or Josephson parametric amplifiers (JPAs)~\cite{admxPRL124} $-$ while there has been an attempt to circumvent it using squeezed-state receivers~\cite{HAYSTAC3}. Yet, it is clear that the turning point in future experiments will be the introduction of single microwave photon counters in the amplification chain~\cite{Lamoreaux,Kuzmin}.

In this work we describe the operation of a classical haloscope of the QUAX--$a\gamma$ experiment using a copper cavity coupled to a JPA and immersed in a static magnetic field of $8.1~\text{T}$, all cooled down with a dilution refrigerator at a working temperature $T \sim 150~\text{mK}$. These features improve the work of Ref.~\cite{QUAX3}, allowing us to exclude values of $g_{a\gamma\gamma}>0.766\times10^{-13}$~GeV$^{-1}$ at 90\% C.L. in a small region of 3.7~neV around $m_a=43.0182~\mu$eV.

In Sec.~\ref{sec:setup} we describe the experimental setup along with its calibration, while in Sec.~\ref{sec:results} we present the results and data analysis. We give prospects for QUAX--$a\gamma$ in Sec.~\ref{sec:conclusions}.

\section{\label{sec:setup}Experimental Setup}

\begin{figure}[h!]
  \centering
      \includegraphics[angle=-90, width=0.47\textwidth]{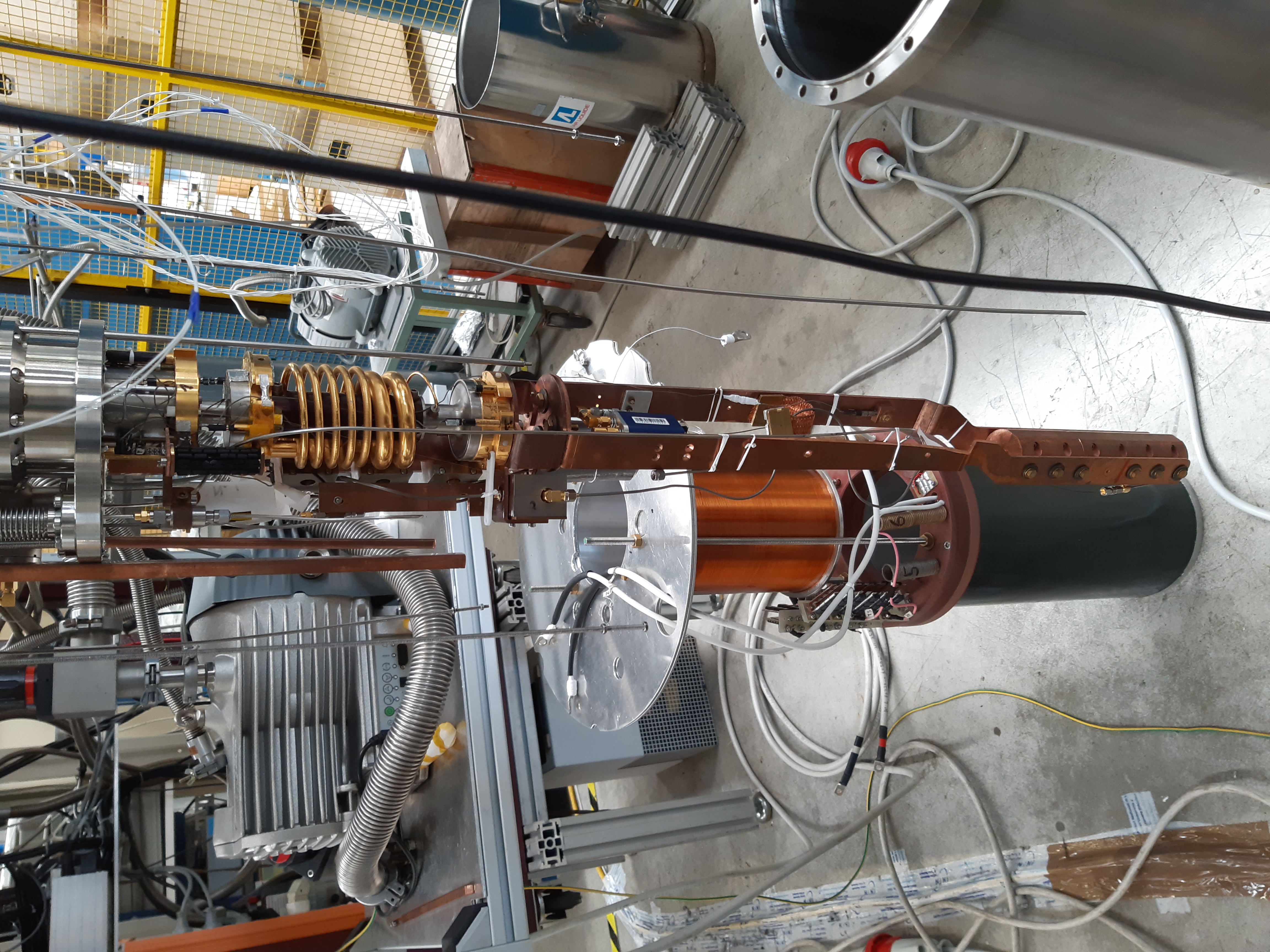}
\caption{\small View of the QUAX$-a\gamma$ dilution refrigerator insert, instrumented with the resonant cavity (at the bottom) and amplification chain. In the background, the 8.1~T magnet with its countercoil is visible.}
\label{fig:cavitaCu}
\end{figure}

The haloscope, assembled at Laboratori Nazionali di Legnaro (LNL), is composed of a cylindrical OFHC-Cu cavity (Fig.~\ref{fig:cavitaCu}), with an inner radius of 11.05~mm and length 210~mm, inserted inside the 150~mm diameter bore of an 8.1~T superconducting (SC) magnet of length 500~mm. The total volume of the cavity is $V = 80.56~\text{cm}^3$. The whole system is hosted in a dilution refrigerator with a base temperature of 90~mK. Each cavity endplate hosts a dipole antenna in the holes drilled on the cavity axis. The cavity is treated with electrochemical polishing to minimize surface losses. We measure the resonant peak of the $TM010$ mode at 150~mK and with the magnet on, with a vector network analyzer, obtaining the frequency $\nu_c$= 10.4018~GHz and an unloaded quality factor $Q_0$= 76,000, in agreement with expectations from simulation performed with the ANSYS HFSS suite~\cite{hfss}. During data-taking runs, the cavity is critically coupled to the output rf line and the loaded quality factor is measured to be about $Q_L$= 36,000.

\begin{figure}[h!]
  \centering
      \includegraphics[width=0.4\textwidth]{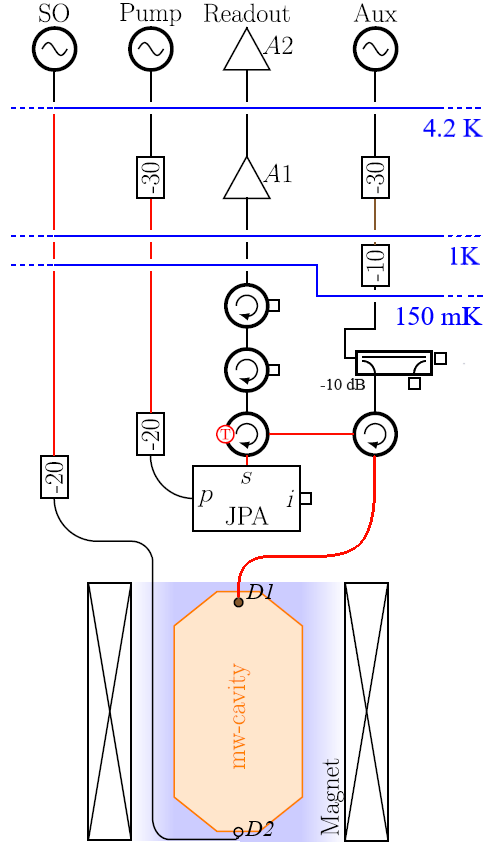}
\caption{\small Schematics of the experimental apparatus. The microwave cavity (orange) is immersed in the uniform magnetic field (blue shaded region) generated by the magnet (crossed boxes). A1 and A2 are the cryogenic and room-temperature amplifiers, respectively. The JPA amplifier has three ports: signal (s), idler (i), and pump (p). Superconducting cables (red) are used as transmission lines for rf signals from the 4~K stage to the 150~mK stage.
Thermometers (red circled T) are in thermal contact with the resonant cavity and the signal port on the JPA.
Attenuators are shown with their reduction factor in decibels. The horizontal lines (blue) identify the boundaries of the cryogenic stages of the apparatus, with the cavity enclosed within the 150~mK radiation shield. The magnet is immersed in liquid helium.}
\label{fig:Apparatus}
\end{figure}

The rf setup is the same as our previous measurement~\cite{QUAX4} and is shown in Fig.~\ref{fig:Apparatus}. It consists of four rf lines used to characterize and measure the cavity signal and to determine attenuations and gains. Starting from the left of Fig.~\ref{fig:Apparatus}, the ``SO'' line connects the source oscillator to the fixed, weakly coupled antenna D2 and is used to inject calibration and probe signals into the cavity. The ``Pump'' line connects the pump-signal generator to the corresponding ``p'' port of a JPA amplifier. The cavity is critically coupled to the ``Readout'' line through antenna D1, tunable via a micrometric screw. The emitted power enters the JPA on the ``s'' port and is reflected, amplified, toward the HEMT cryogenic (A1) and HEMT room-temperature (A2) amplifiers. The signal is then downconverted with an I-Q mixer with a 100~MHz IF band, the phase and quadrature components of the heterodyne signal are postamplified in a 10 MHz band and finally  sampled via an analog-to-digital converter (ADC) with a 2~MHz bandwidth. The ``Aux'' line is an auxiliary line introduced for calibration purposes. To minimize the Johnson noise contribution at the coldest stage, we insert attenuators and circulators in the rf lines. A nonoptimal attenuation of the ``Aux'' line with 10 dB attenuation at 1~K and 10 dB at 150~mK causes an excess Johnson noise of about 95~mK on the circulator and on the cavity (since they are thermally connected), corresponding to an effective temperature of the circulator of 273~mK at 10 GHz.
We monitor the temperatures with RuO$_2$ thermometers, one in thermal contact with the cavity and the other with the mixing chamber. Due to some unexpected behavior of the thermometers, we only estimate a temperature between 100~mK and 150~mK in the mixing chamber and between 200~mK and 250~mK on the cavity.

The JPA in our setup, first realized in~\cite{Roch}, has noise temperature expected at the quantum limit of about 0.5~K (including 0.25~K from vacuum fluctuations) and a resonance frequency tunable between 10 and 10.5 GHz by varying the pump amplitude and frequency and by applying a small magnetic field for fine regulation. After tuning the resonance frequency of the JPA to that of the cavity we measure, a gain of 18~dB in a 10~MHz bandwidth.

\section{\label{sec:results}Analysis and Results}

We first measure the transmittivity of the rf lines and the amplification gain as described in detail in~\cite{QUAX4}. Then we calibrate the power scale by injecting a known signal. Finally, we measure the system nois temperature resulting in $T_n=(0.99\pm0.15_{cal}\pm0.04_{stab})$~K, where the errors result from the uncertainty in the calibration scale due to a limited tunability of the coupling of antenna D1 and to the temperature variation during the data-taking run. This value, within the error, is in agreement with our estimate of 0.83~K obtained from the single contributions reported in Table~\ref{tab:noise}.

\begin{figure}[tb]
  \centering
      \includegraphics[width=0.47\textwidth]{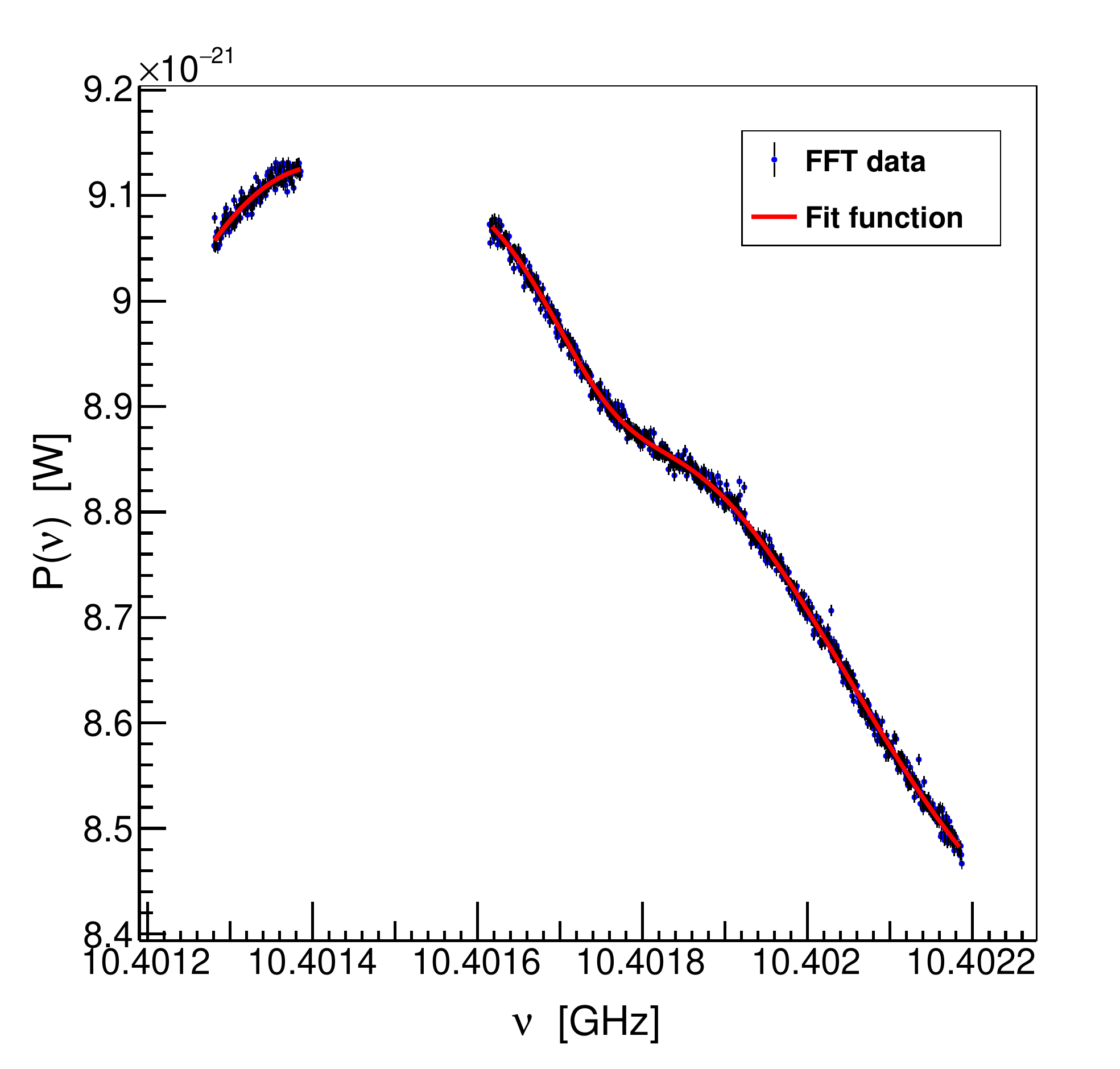}
\caption{\small Measured power spectrum. The red line represents the model function used to obtain residuals.}
\label{fig:Fit}
\end{figure}

\begin{table}
  \begin{center}
    \caption{Noise contributions estimated at the cavity resonance frequency. ``Vacuum" is the contribution of quantum fluctuations of vacuum. The room-temperature HEMT (A2) contribution is negligible. ``Cables" refers to rf attenuation of cables, with the only effect that of reducing the overall gain.}
    \label{tab:noise}
  \vspace*{0.5cm}
    \begin{tabular}{c|c|c|c}
			\hline\hline
      Source & Gain [dB] &  Noise temp. [K] & Input noise [K] \\\hline
      Cavity & -- & 0.078 & 0.078 \\
			Vacuum & -- & 0.25  & 0.25 \\
			JPA & 18 & 0.25 & 0.25 \\
			Cables & -3 & -- & -- \\
			HEMT (A1)  & 30 & 8 & 0.25 \\
			\hline
			\multicolumn{3}{c|}{Total} & 0.83 \\
			\hline
    \end{tabular}
  \end{center}
\end{table}
After setting the magnetic field to 8.1~T, we perform the axion search for a total time $\Delta t =4203$~s with an ADC sampling of 2~Ms/s with the cavity tuned at a fixed-frequency of $\nu_c=10.4018$~GHz. We compute the average power spectrum with a fine frequency bin of 651~Hz, corresponding to 1/16th of the expected axion-signal width~\cite{Turner}; we then identify and remove IF noise bins, which have a width $\Delta \nu_{\scriptscriptstyle \textup{IF}} << \Delta \nu_{\textup{bin}}$. We exclude from our analysis a 200~kHz frequency region around the local oscillator frequency, $\nu_{lo}=10.4015$~GHz, which is affected by $1/f$ and pickup noise, also appearing when running the setup with the magnet off. For the same reason we also exclude a single bin in the cavity region; this has an off-resonance counterpart, symmetric with respect to the local oscillator. Performing the ratio of the left half of the spectrum to the right half, the single bin and its counterpart perfectly cancel; thus they are considered noise bins and are removed. This single bin and the 200 kHz region around LO are the only features removed from the spectrum. Finally, we considere only the region of the Lorentzian distribution of the cavity power spectrum with an expected power of at least 10\% of the peak value. The resulting spectrum is shown in Fig.~\ref{fig:Fit}.

In order to extract the residuals, we model the system composed of the cavity and the ``Readout'' line with an equivalent electrical circuit. Using the transmission-line formalism, we derive the following expression of the power spectrum:
\begin{widetext}
\begin{equation}
  P_{n}=G_{\scriptscriptstyle TOT}(\omega) k_B (\tilde{T}_1+T_{A,tot}) \left( \frac{\tilde{T}_1}{T_{A,tot}+\tilde{T}_{1}} \frac{\tilde{T}_c/\tilde{T}_1  +(Q_L\delta)^2}{1+(Q_L\delta)^2} + \frac{T_{A,tot}}{T_{A,tot}+\tilde{T}_1} \right).
\end{equation}
\end{widetext}
Note that $T_{A,tot}=0.50$~K is the sum of the noise temperatures of the JPA and HEMT (A1) amplifiers, as reported in Table~\ref{tab:noise}. Here, $T_1\sim 270$~mK is the effective temperature of the circulator on the ``Aux" line, and $T_c$ is the temperature of the cavity, which is left as a free parameter. Low temperatures require the use of Bose-Einstein distribution, so instead of $T_1$ and $T_c$ we use the noise temperatures $\tilde{T}_1$ and $\tilde{T}_c$; the tilde stands for $k_{B} \tilde{T}=h\nu_{c}/(\exp{(h\nu_c/k_BT)}-1)+h\nu_c/2$, including the contribution from vacuum fluctuations. Therefore, the first term in the big parentheses represents the contribution of the circulator's Johnson noise reflected by the cavity and the thermal noise emitted by the cavity itself, while the second term is the added noise of amplifiers.
Here, $Q_L$ is the loaded quality factor, $\delta=(\nu/\nu_{c}-\nu_c/\nu)$, $\nu_c$ the cavity resonance frequency and $G_{\scriptscriptstyle TOT}(\omega)$ is the total gain function.\newline
\indent We fit the power spectrum by expressing $G_{\scriptscriptstyle TOT}(\omega)$ as 2nd and 4th order polynomials in the left and right branches, respectively. Given the large number of unknown parameters we fix all known quantities to the best of our knowledge, taking into account measurement errors. The best fit is obtained for $\nu_c=10.40176$~GHz
and $Q_L=35,000$, in reasonable agreement with our measurements. When fixing the ``Aux" circulator temperature to $T_1=273$~mK, we obtain a cavity temperature $T_c=250$~mK, compatible with our expectations. The fit has $\chi^2/n=1226/1032$ and is shown by the red line in Fig.~\ref{fig:Fit}. Changing $T_1$ in an interval between 150 and 273~mK does not impact the quality of the fit and just reduces the cavity temperature down to about 100~mK in the former case.

\begin{figure}[h!]
  \centering
      \includegraphics[width=0.47\textwidth]{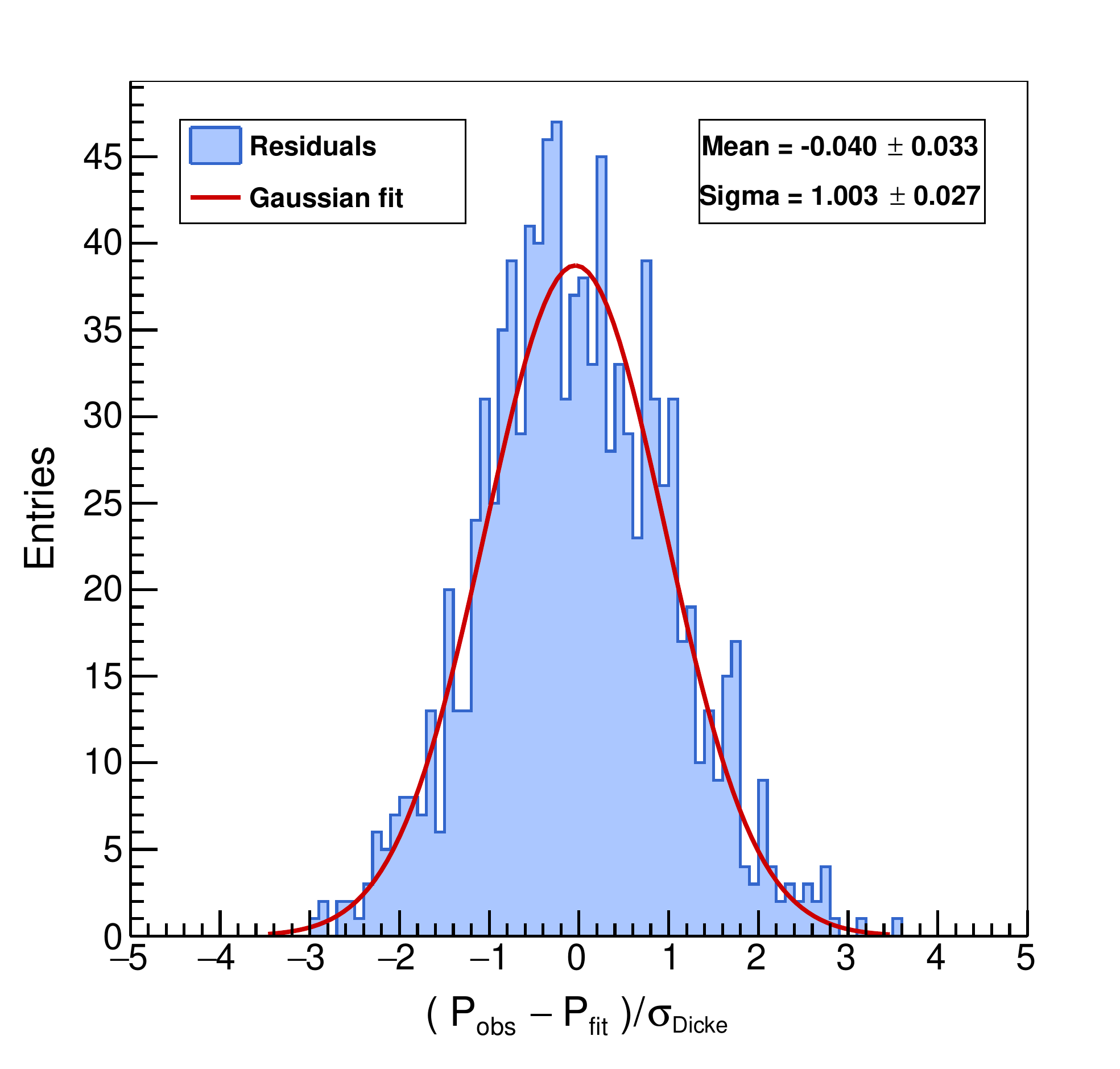}
\caption{\small Distribution of the residuals normalized to the expected thermal noise.}
\label{fig:ResidualFit}
\end{figure}

Since the expected axion signal width is of about 10~kHz in the lab frame~\cite{Sikivie2,Turner}, with a bin width of 651~Hz a power excess is expected in about 16 consecutive bins. We normalize the residuals obtained in the fit procedure to the expected noise power $\sigma_{\scriptscriptstyle \textup{Dicke}}=5.38\times 10^{-24}$~W derived from the Dicke radiometer formula~\cite{Dicke} using the system temperature $T_n=0.99$~K
\begin{equation}
  \sigma_{\scriptscriptstyle \textup{Dicke}} = k_B T \sqrt{\Delta \nu/\Delta t}\, ,
\end{equation}
where $\Delta t$ is the integration time.
The distribution of the normalized residuals is shown in Fig.~\ref{fig:ResidualFit} together with the result of a Gaussian fit, showing a rms compatible with 1.

\begin{figure}[h!]
  \centering
      \includegraphics[width=0.47\textwidth]{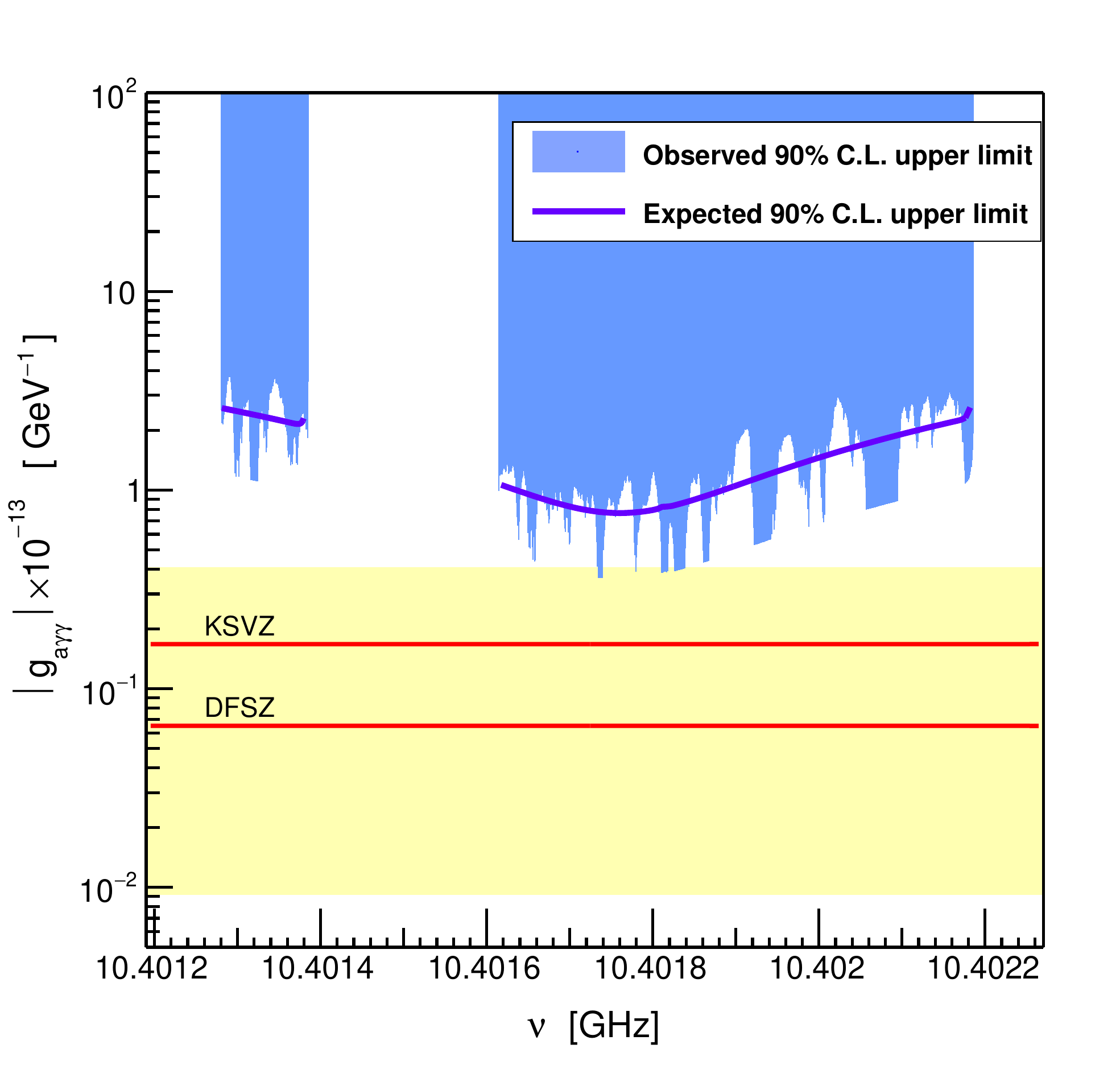}
\caption{\small The 90\% single-sided C.L. upper limit for the axion coupling constant $g_{a \gamma \gamma}$. Each point corresponds to a test axion mass in the analysis window. The solid curve represents the expected limit in the case of no signal. The yellow region indicates the QCD axion model band. We assume $\rho_a\sim0.45$~GeV/cm$^3$.}
\label{fig:Bin90limit}
\end{figure}

To claim a discovery candidate we require that power is in excess of 5$\sigma$ from the noise spectrum, corresponding to some bins $>5$ in the normalized residuals. Correcting for the look-elsewhere effect, the requirement would be to find excesses greater than $Z=\Phi^{-1} \left( 1- 2.87\times 10^{-7}/N_{\text{bin}} \right)$, where $\Phi$ is the cumulative of the normal distribution and $N_{\text{bin}}=1041$ is the number of data bins, corresponding to an effective number of $Z=6.204$.
We did not find any candidate, so we interpret our result as an exclusion test for the axion existence in this mass range.
A maximum likelihood approach is used to compute the estimator $\hat{g}_{a\gamma\gamma}$ from the data, with the logarithmic likelihood
\begin{equation}
  -2 \ln \mathcal{L}(m_a, g_{a\gamma\gamma}^2) = \sum_{i=0}^{N_{\text{bin}}} \frac{\left( R_i - S_i\left( m_a, g_{a\gamma\gamma}^2 \right) \right)^2 }{(\sigma_{\scriptscriptstyle \textup{Dicke}}^{\scriptscriptstyle \textup{max}})^2} \, ,
\label{eq:ML}
\end{equation}
where we have assumed Gaussian statistics. The index $i$ runs over bins, $R_i$ are the observed residuals, $S_i$ are the model signals given by Eq.~(\ref{eq:power}) multiplied by Eq.~(\ref{eq:lorentz}) and convoluted with the full Standard Halo Model distribution~\cite{Turner}, and $\sigma_{\scriptscriptstyle \textup{Dicke}}^{\scriptscriptstyle \textup{max}}=6.41\times 10^{-24}$~W is the most conservative noise power, obtained with the maximum temperature allowed within its error. Note that the rhs of Eq.~(\ref{eq:ML}) is a $\chi^2$ distribution. The estimator $\hat{g}^2_{a\gamma\gamma}$ is then evaluated by solving $\partial \chi^2 / \partial g^2_{a\gamma\gamma} = 0$, and its error is calculated as $\sigma_{\hat{g}^2_{a\gamma\gamma}}=\left(1/2 \, \partial^2 \chi^2 / (\partial g^2_{a\gamma\gamma})^2 \right)^{-1/2}$. The maximum likelihood procedure is repeated for each axion test mass, precisely $N_{\text{bin}}$ times, resulting in a step size of 651~Hz.
Finally, we calculate the limit to the axion-photon coupling with a 90\% confidence level,  power-constraining values of $\hat{g}^2_{a\gamma\gamma}$ that underfluctuate below $-1\sigma$~\cite{Cowan}.
We show in Fig.~\ref{fig:Bin90limit} the limit as a function of the tested axion masses, shown with a colored area, together with a solid purple line showing the expected limit in the case of no signal. The reference upper limit of our search is the value at the maximum sensitivity (the minimum of the purple line of Fig.~\ref{fig:Bin90limit}), $g_{a\gamma\gamma}^{\scriptscriptstyle \textup{CL}} < 0.766\times10^{-13}~\textup{GeV}^{-1}$ at 90\% C.L.

\begin{figure}[h!]
  \centering
      \includegraphics[width=0.47\textwidth]{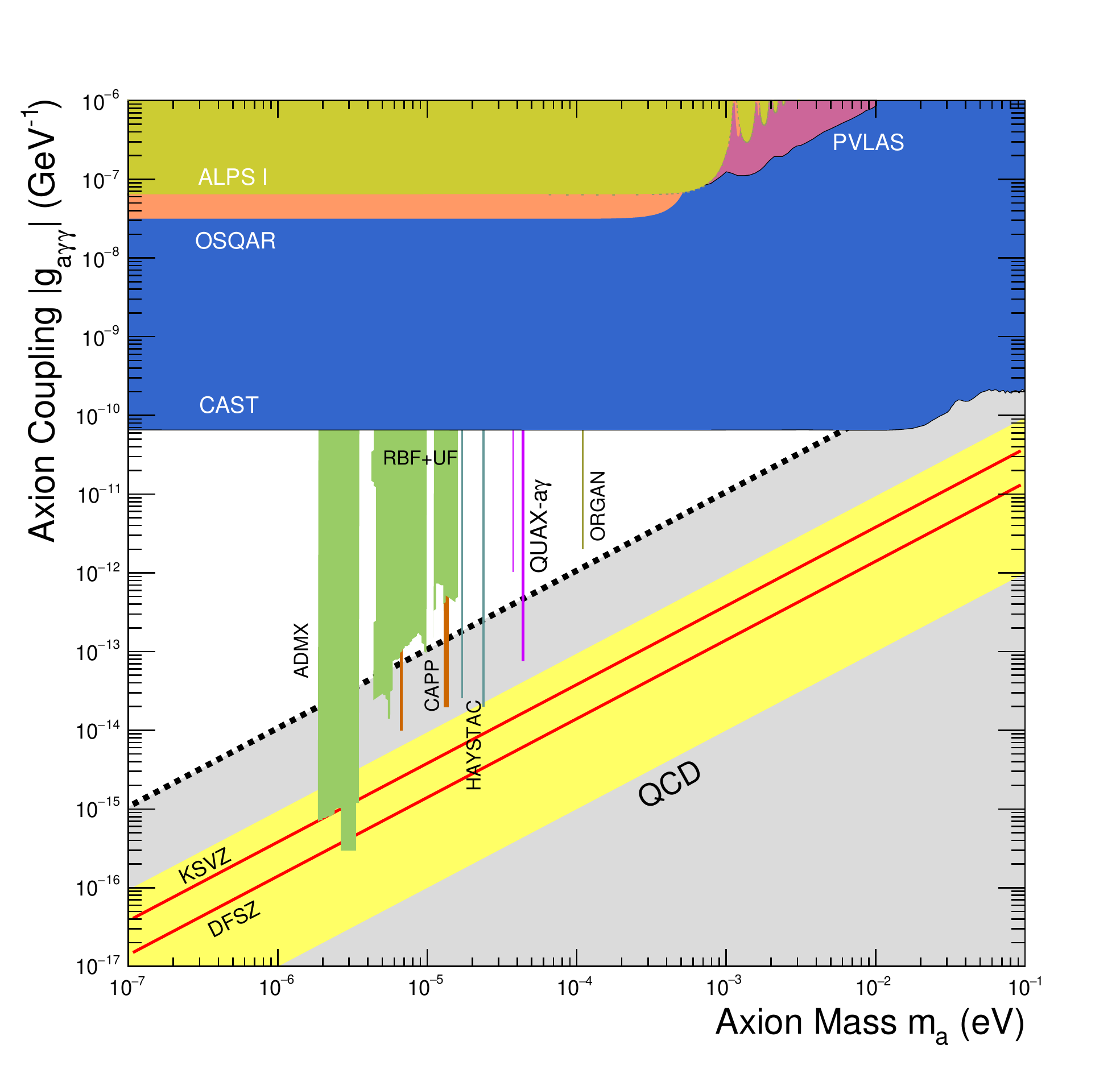}
\caption{\small Aggregate plot of the limits on $g_{a\gamma\gamma}$ obtained from the main axion search experiments. The two limits obtained by the QUAX Collaboration are highlighted in purple. The gray area identifies the region where axions could be found, with the yellow band and the two solid red lines identifying the coupling predicted by the KSVZ and DFSZ models and its uncertainty.}
\label{fig:ExclusionPlot}
\end{figure}

In Fig.~\ref{fig:ExclusionPlot} we compare the limit $g_{a\gamma\gamma}^{\scriptscriptstyle \textup{CL}}$ that we observed, in a mass window $\Delta m_a=3.7$~neV centered at the mass $m_a=43.0182~\mu$eV, with those obtained in previous searches.

\section{\label{sec:conclusions}Conclusions}
We report results of the search with a haloscope for galactic axions with mass of about 43~$\mu$eV in a small frequency region of 3.7~neV. By cooling the system to about 150~mK in a dilution refrigerator and employing a Josephson parametric amplifier with noise at the standard quantum limit, we set a limit on the axion-photon coupling of about a factor 2 from the QCD band. We show directly that, even at a frequency as large as 10 GHz, haloscopes will soon have the sensitivity to observe QCD axions. The total noise, estimated as twice the standard quantum limit, can be further reduced by improving the thermalization of the resonant cavity and the line filtering, and by reducing the noise contribution from the HEMT.

\begin{acknowledgments}
We are grateful to E. Berto, A. Benato, and M. Rebeschini for the mechanical work; F. Calaon and M. Tessaro for help with the electronics and cryogenics, and  to F. Stivanello for the chemical treatments. We thank G. Galet and L. Castellani for the development of the magnet power supply, and M. Zago who realized the technical drawings of the system. We deeply acknowledge the Cryogenic Service of the Laboratori Nazionali di Legnaro for providing us with large quantities of liquid helium on demand.
\end{acknowledgments}

\bibliography{biblioFile}


\end{document}

%% file: authors.tex
\author{D.~Alesini} \affiliation{INFN, Laboratori Nazionali di Frascati, Frascati, Roma, Italy}
\author{C.~Braggio} \affiliation{INFN, Sezione di Padova, Padova, Italy} \affiliation{Dipartimento di Fisica e Astronomia, Padova, Italy}
\author{G.~Carugno} \affiliation{INFN, Sezione di Padova, Padova, Italy} \affiliation{Dipartimento di Fisica e Astronomia, Padova, Italy}
\author{N.~Crescini} \altaffiliation{Present address: IBM Research-Z{\"u}rich, S{\"a}umerstrasse 4, CH-8803 R{\"u}schlikon, Switzerland} \affiliation{INFN, Laboratori Nazionali di Legnaro, Legnaro, Padova, Italy} \affiliation{Dipartimento di Fisica e Astronomia, Padova, Italy}
\author{D.~D'Agostino} \affiliation{Dipartimento di Fisica E.R. Caianiello, Fisciano, Salerno, Italy} \affiliation{INFN, Sezione di Napoli, Napoli, Italy}
\author{D.~Di~Gioacchino} \affiliation{INFN, Laboratori Nazionali di Frascati, Frascati, Roma, Italy}
\author{R.~Di Vora}\email{divora@pd.infn.it} \affiliation{INFN, Sezione di Padova, Padova, Italy} \affiliation{Dipartimento di Scienze Fisiche, della Terra e dell'Ambiente, Universi{\`a} di Siena, via Roma 56, 53100 Siena, Italy}
\author{P.~Falferi} \affiliation{Istituto di Fotonica e Nanotecnologie, CNR Fondazione Bruno Kessler, I-38123 Povo, Trento, Italy} \affiliation{INFN, TIFPA, Povo, Trento, Italy}
\author{U.~Gambardella} \affiliation{Dipartimento di Fisica E.R. Caianiello, Fisciano, Salerno, Italy} \affiliation{INFN, Sezione di Napoli, Napoli, Italy}
\author{C.~Gatti} \affiliation{INFN, Laboratori Nazionali di Frascati, Frascati, Roma, Italy}
\author{G.~Iannone} \affiliation{Dipartimento di Fisica E.R. Caianiello, Fisciano, Salerno, Italy} \affiliation{INFN, Sezione di Napoli, Napoli, Italy}
\author{C.~Ligi} \affiliation{INFN, Laboratori Nazionali di Frascati, Frascati, Roma, Italy}
\author{A.~Lombardi} \affiliation{INFN, Laboratori Nazionali di Legnaro, Legnaro, Padova, Italy}
\author{G.~Maccarrone} \affiliation{INFN, Laboratori Nazionali di Frascati, Frascati, Roma, Italy}
\author{A.~Ortolan} \affiliation{INFN, Laboratori Nazionali di Legnaro, Legnaro, Padova, Italy}
\author{R.~Pengo} \affiliation{INFN, Laboratori Nazionali di Legnaro, Legnaro, Padova, Italy}
\author{A.~Rettaroli}\email{alessio.rettaroli@lnf.infn.it} \affiliation{INFN, Laboratori Nazionali di Frascati, Frascati, Roma, Italy} \affiliation{Dipartimento di Matematica e Fisica, Universit{\`a} di Roma Tre, Roma, Italy}
\author{G.~Ruoso} \affiliation{INFN, Laboratori Nazionali di Legnaro, Legnaro, Padova, Italy}
\author{L.~Taffarello} \affiliation{INFN, Sezione di Padova, Padova, Italy}
\author{S.~Tocci} \affiliation{INFN, Laboratori Nazionali di Frascati, Frascati, Roma, Italy}